\documentclass[conference]{IEEEtran}
\IEEEoverridecommandlockouts
\usepackage{cite}
\usepackage{amsmath,amssymb,amsfonts}
\usepackage{algorithmic}
\usepackage{graphicx}
\usepackage{textcomp}
\usepackage{xcolor}
\def\BibTeX{{\rm B\kern-.05em{\sc i\kern-.025em b}\kern-.08em
    T\kern-.1667em\lower.7ex\hbox{E}\kern-.125emX}}
\begin{document}

\title{Analyzing programming languages by community characteristics on Github and StackOverflow}

\author{
\IEEEauthorblockN{Samarth Tambad}
\IEEEauthorblockA{
\textit{Courant Institute of Mathematical} \\
\textit{Sciences} \\
\textit{New York University}\\
New York, NY, USA \\
svt258@nyu.edu}
\and
\IEEEauthorblockN{Rohit Nandwani}
\IEEEauthorblockA{
\textit{Courant Institute of Mathematical} \\
\textit{Sciences} \\
\textit{New York University}\\
New York, NY, USA \\
rhn235@nyu.edu}
\and
\IEEEauthorblockN{Suzanne K. McIntosh}
\IEEEauthorblockA{
\textit{Courant Institute of Mathematical} \\
\textit{Sciences} \\
\textit{Center for Data Science} \\
\textit{New York University}\\
New York, NY, USA \\
mcintosh@cs.nyu.edu }
}

\maketitle

\begin{abstract}
The choice of programming language is a very important decision as it not only affects the performance and maintainability of the software but also dictates the talent pool and community support available. To better understand the trade-offs involved in making such a decision, we define and compute popularity, demand, availability and community engagement of programming languages through online collaboration platforms. We perform our analysis using data from Github and StackOverflow, two of the most popular programming communities. We get data related projects, languages and developer engagement from Github and programming questions with answers along with language tags from StackOverflow. We compute metrics separately for the two data sources and then combine the metrics to provide a holistic and robust picture of the communities for the most popular programming languages.
\end{abstract}


\section{Introduction}
Software development and maintenance is a complex activity involving many important decisions that need to be made. The choice of programming language is one such decision. From the perspective of the managers of the software projects, this decision not only affects the performance of the software but also dictates the talent pool and community support available. From the perspective of the developer, it dictates the current job opportunities and their future career trajectory. 

We analyse the popularity and the community friendliness of programming languages and estimate the availability and demand of developers proficient in them. For our analysis, we look at data from Github and StackOverflow, two of the most popular programming communities. 

Github is a platform for collaborative software development. Data gathered from this platform is suitable for measuring the popularity of languages and availability/demand of developers. Particularly the information available about repositories such as languages used and contributions made by developers is useful.

StackOverflow is a popular online programming Q\&A community providing its participants with rapid access to knowledge and expertise of their peers. The community support is a valuable tool for developers in any programming language. Therefore, a more open, welcoming and  responsive (i.e. friendly) community is a good thing to have in order to be more productive as a developer. Data such as the questions asked and the quality and time-frame of the response is a good indicator of the “friendliness” of a particular programming community.

We combine the data gathered from the two sources to compute the metrics of popularity, community engagement, availability and demand \footnote{Code repository is at https://github.com/samarthtambad/big-data-pl}. These metrics provide a holistic view of the pros/cons of different languages. We then use these metrics to compare different languages and help answer questions such as: which is the first language I should learn?, which language is most in demand right now?, suggest an alternative language because I work with x language but the community support is bad, etc. 

The remainder of this paper is organised as follows: we describe our motivation in Section II followed by a survey of related work in Section III. In Section IV, we provide a detailed description of the datasets used. We describe our analytic in Section V followed by application design in Section VI. In Section VII we describe our experimental setup and analysis of results. In Section VIII we provide our conclusions and provide scope for future work in Section IX.

\section{Motivation}
Open source has been gaining popularity among the developer community. Increasingly, many companies are also realising the benefit of contributing to open source projects which may benefit their business directly or indirectly \cite{lerner2005economics}. Also, developers are increasingly realising the benefit of contributing to open source. Therefore, analysis on the open source developer community is good proxy for the developer community in general.

While choosing a programming language to learn or build a project, it is important to understand the characteristics and strengths of the landscape of programming languages. At the same time, it is critical to have an active and cooperative community for the programming language under consideration to speed up the learning and building process. We find that there is a lack of research on the latter aspect, which combines data from multiple available sources. 

The choice of language based on community has a massive impact on the levels of productivity for the developer and the company \cite{delorey2007programming}, performance of the applications, and the overall satisfaction of the development process\cite{edwards2014developer}. It will also result in increased demand for the developers in the language with better community characteristics. 

There are multiple studies mapping developer productivity and satisfaction, to the profitability of the company\cite{edwards2014developer}\cite{bhatti2007impact}. Programming languages can also have a major impact in the career trajectories and overall satisfaction of developers.

\section{Related Work}
In this section, we describe the related work compiled from the literature.

\subsection{Analysis using Github data}
Ray, B. et al\cite{ray2014large}, perform a large scale study on the quality of code with respect to programming languages using text mining and regression techniques. They find that there is a significant correlation between the two.

Kalliamvakou et al\cite{kalliamvakou2016depth} describe the perils on mining data on GitHub. They point that inactive account, invisible merges on pull requests, public activity on repositories could cause problems in analysis and how to overcome them.

\subsection{Analysis using StackOverflow data}
Jie Yang et al\cite{yang2014sparrows} study the characteristics of experts on StackOverflow. They give us important metrics such as the debatableness of a question and the utility of an answer. 

Seyed Mehdi Nasehi et al\cite{nasehi2012makes} describe what makes a good code example on StackOverflow by analyzing the interactions with code examples. 

Blerina Bazelli et al\cite{bazelli2013personality} describe the personality traits of successful contributors on StackOverflow including extroversion and negativity.

Gupta, R. et al\cite{gupta2016learning} study reopened questions on StackOverflow, and suggest the editing questions / answers even after acceptance/closing is a good sign of expertise in the community.

Wang, S et al\cite{wang2013empirical} study if the population on StackOverflow can be divided into givers or takers. They also model the types of questions asked using LDA.

\subsection{Analysis using Github and StackOverflow combined}
Lee, R. et al\cite{lee2017github}, compare the developer interests on Github and StackOverflow and suggest a high correlation between the two. This helps us know the differences in proportion of contribution on the two different collaboration platforms.

Badashian, A. S et al\cite{badashian2014involvement} provide methods and metrics to measure core contributions, editorial activities and influence on Github and StackOverflow.

Vasilescu, B. et al\cite{vasilescu2013stackoverflow} show how activity on StackOverflow impacts the activity on Github and vice versa.

Tian, Y et al\cite{tian2019geek} measure the quality of individual contributors on the two platforms and combine the measures across both platforms for each individual.

\section{Datasets}
There are two main datasets used in our analysis. Here we describe the datasets in detail along with their schema.

\subsection{Github}
Github is a collaborative software development platform that allows code sharing and version control. Developers can perform various activities such as creating, forking or committing to a repository, opening issues or submitting pull requests to contribute someone else’s repository. The programming language used is tagged for each repository which is very helpful.

We collected the data from the GH Torrent project \cite{gousios2013ghtorent}. It is a dump of Github usage data over the period ranging from October 2013 to June 2019. The data is separated into multiple tables and stored as csv files. The list of tables is given in Table \ref{gh_tables} and the descriptions for each with their complete schema is given in Tables \ref{gh_projects} - \ref{gh_issue_events}.

\begin{table}[htbp]
\caption{Github tables data}
\label{gh_tables}
\begin{center}
\begin{tabular}{|c|c|c|}
\hline
\textbf{Tables} & \textbf{Schema} & \textbf{\textit{Description}} \\
\hline
projects.csv & Table \ref{gh_projects} & Github project repositories \\
\hline
commits.csv & Table \ref{gh_commits} & A list of all commits on Github. \\
\hline
pull\_requests.csv & Table \ref{gh_pr} & List of pull requests for repos \\
\hline
pull\_request\_history.csv & Table \ref{gh_pr_history} & Chronologically ordered \\
 & & list of events on a pull request  \\
\hline
issues.csv & Table \ref{gh_issues} & Issues that have been recorded \\
 & & for a project \\
\hline
issue\_events.csv & Table \ref{gh_issue_events} & Chronologically ordered \\
 & & list of events on an issue \\
\hline
\end{tabular}
\end{center}
\end{table}

\begin{table}[htbp]
\caption{Projects: $138205530$ rows}
\label{gh_projects}
\begin{center}
\begin{tabular}{|c|c|c|c|c|}
\hline
\textbf{Columns} & \textbf{\textit{Data Type}}& \textbf{\textit{Min}}& \textbf{\textit{Max}} & \textbf{\textit{Distinct}} \\
\hline
id & Integer & 1 & 137611262 & 65296722 \\
\hline
owner\_id & Integer & 1 & 51697268 & 10778579 \\
\hline
language & String & 1 & 24 & 376 \\
\hline
year & Integer & 2007 & 2019 & 13 \\
\hline
\end{tabular}
\label{tab1}
\end{center}
\end{table}

\begin{table}[htbp]
\caption{Commits: $1353856359$ rows}
\label{gh_commits}
\begin{center}
\begin{tabular}{|c|c|c|c|c|}
\hline
\textbf{Columns} & \textbf{\textit{Data Type}}& \textbf{\textit{Min}}& \textbf{\textit{Max}} & \textbf{\textit{Distinct}} \\
\hline
id & Integer & 1 & 1415397637 & 1353856359 \\
\hline
author\_id & Integer & 1 & 51697270 & 21406056 \\
\hline
committer\_id & Integer & 1 & 51697269 & 18732264 \\
\hline
project\_id & Integer & 1 & 137611262 & 73223832 \\
\hline
year & Integer & 2007 & 2019 & 13 \\
\hline
\end{tabular}
\label{tab1}
\end{center}
\end{table}

\begin{table}[htbp]
\caption{Pull Requests: $51730295$ rows}
\label{gh_pr}
\begin{center}
\begin{tabular}{|c|c|c|c|c|}
\hline
\textbf{Columns} & \textbf{\textit{Data Type}}& \textbf{\textit{Min}}& \textbf{\textit{Max}} & \textbf{\textit{Distinct}} \\
\hline
id & Integer & 6350 & 64121451 & 51730295 \\
\hline
head\_repo\_id & Integer & 3 & 137611190 & 13435966 \\
\hline
base\_repo\_id & Integer & 2 & 137611003 & 7099278 \\
\hline
head\_commit\_id & Integer & 23 & 1415397634 & 50517219 \\
\hline
base\_commit\_id & Integer & 14 & 1415395953 & 36332904 \\
\hline
pull\_request\_id & Integer & 1 & 231471 & 189365 \\
\hline
\end{tabular}
\label{tab1}
\end{center}
\end{table}

\begin{table}[htbp]
\caption{Pull Request History: $134649096$ rows}
\label{gh_pr_history}
\begin{center}
\begin{tabular}{|c|c|c|c|c|}
\hline
\textbf{Columns} & \textbf{\textit{Data Type}}& \textbf{\textit{Min}}& \textbf{\textit{Max}} & \textbf{\textit{Distinct}} \\
\hline
id & Integer & 15017368 & 152227176 & 134649096 \\
\hline
pull\_request\_id & Integer & 6350 & 64121451 & 51884523 \\
\hline
action & String & 6 & 11 & 6 \\
\hline
actor\_id & Integer & 1 & 51697114 & 4484020 \\
\hline
year & Integer & 2010 & 2019 & 10 \\
\hline
\end{tabular}
\label{tab1}
\end{center}
\end{table}

\begin{table}[htbp]
\caption{Issues: $98076172$ rows}
\label{gh_issues}
\begin{center}
\begin{tabular}{|c|c|c|c|c|}
\hline
\textbf{Columns} & \textbf{\textit{Data Type}}& \textbf{\textit{Min}}& \textbf{\textit{Max}} & \textbf{\textit{Distinct}} \\
\hline
id & Integer & 2 & 110037555 & 98076172 \\
\hline
repo\_id & Integer & 1 & 137611003 & 9498704 \\
\hline
issue\_id & Integer & 0 & 337847 & 295187 \\
\hline
year & Integer & 1970 & 2019 & 22 \\
\hline
\end{tabular}
\label{tab1}
\end{center}
\end{table}

\begin{table}[htbp]
\caption{Issue Events: $136108876$ rows}
\label{gh_issue_events}
\begin{center}
\begin{tabular}{|c|c|c|c|c|}
\hline
\textbf{Columns} & \textbf{\textit{Data Type}}& \textbf{\textit{Min}}& \textbf{\textit{Max}} & \textbf{\textit{Distinct}} \\
\hline
event\_id & Integer & 2 & 2147483633 & 124838780 \\
\hline
issue\_id & Integer & 2 & 110035665 & 31533578 \\
\hline
action & String & 6 & 24 & 35 \\
\hline
year & Integer & 1999 & 2019 & 21 \\
\hline
\end{tabular}
\label{tab1}
\end{center}
\end{table}

\subsection{StackOverflow}
StackOverflow is the largest peer reviewed Q/A system for computer programming. All the data is open source and it is available here. It has data from the year 2008 to 2020. We are primarily interested in the 'Posts' dataset which is all the questions, answers and the interactions with them. There are  more than 15M posts with total size of 15 GB with over 500 programming languages. We filter the dataset top 50 most frequent programming languages on Github, narrowing the dataset to 12M posts. The dataset also has a score field which is the sum of upvotes and follows ranging from -146 to 24245 for the posts. We narrow our analysis to the columns listed in Table \ref{so_posts}.

\begin{table}[htbp]
\caption{Posts: $13851898$ rows}
\label{so_posts}
\begin{center}
\begin{tabular}{|c|c|c|c|c|c|}
\hline
\textbf{Columns} & \textbf{\textit{Data Type}}& \textbf{\textit{Min}}& \textbf{\textit{Max}} & \textbf{\textit{Distinct}} \\
\hline
\_Id & Integer & 4 & 60472846 & 11746372 \\
\hline
\_OwnerUserId & Integer & 1 & 12987310 & 2734937 \\
\hline
\_PostTypeId & Integer & 1 & 1 & 1 \\
\hline
\_Score & Integer & -146 & 24124 & 1591 \\
\hline
\_Tag & String & 1 & 16 & 270 \\
\hline
\_CreationYear & Integer & 2008 & 2020 & 13 \\
\hline
\_AnswerCount & Integer & 0 & 296 & 89 \\
\hline
\end{tabular}
\label{tab1}
\end{center}
\end{table}

\begin{figure*}[ht]
\caption{Architecture design diagram}
\label{fig:architecture}
\centerline{%
\includegraphics[width=0.85\textwidth]{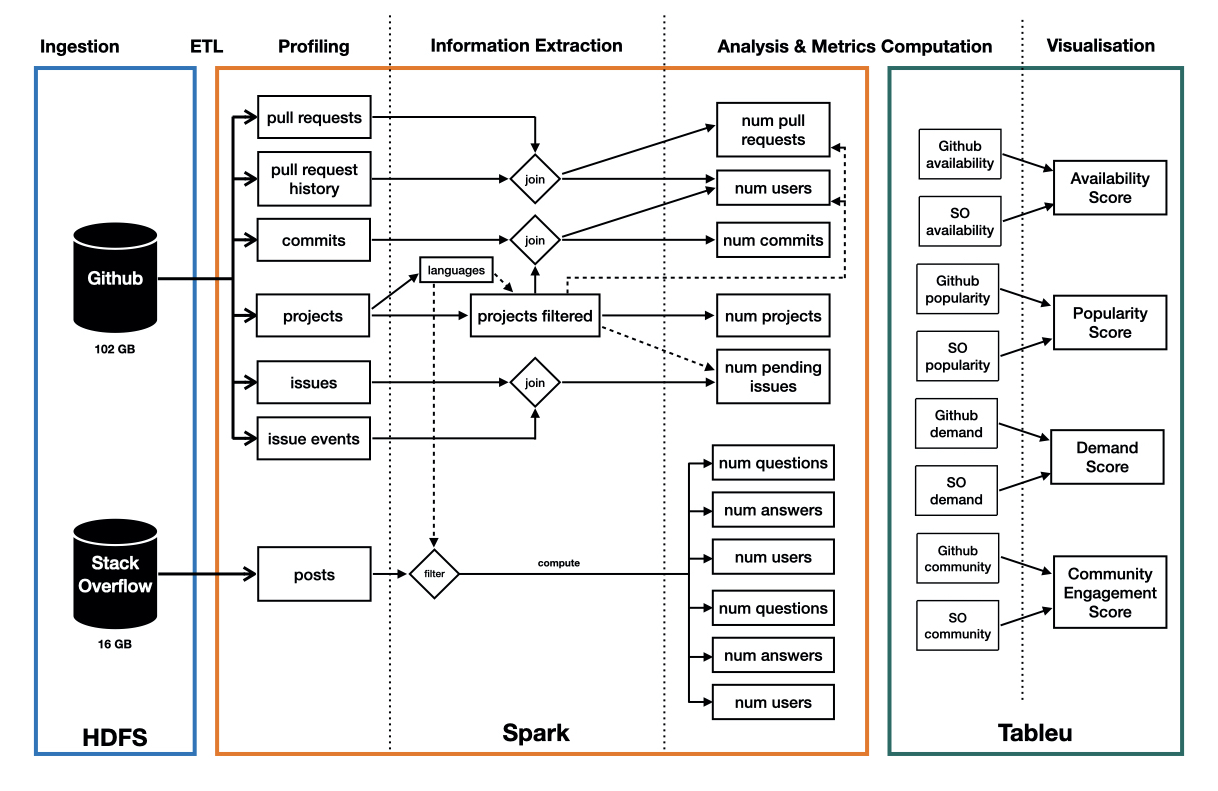}
}
\end{figure*}

\section{Description of Analytic}
In this section, we describe our analytic derived from each source and how we combine it.

\subsection{Github}
With the Github data, we used the relevant tables listed and described in Table \ref{gh_tables} to compute the following intermediate metrics:
\begin{itemize}
  \item \textbf{num\_users - } total number of new users each year, grouped by associated language.  
  \item \textbf{num\_projects - } total number of new projects created each year, grouped by associated language.
  \item \textbf{num\_commits - } total number of commits made to repositories each year, grouped by associated language.
  \item \textbf{num\_pull\_requests - } total number of pull requests opened to base repositories each year, grouped by associated language.
  \item \textbf{num\_pending\_issues - } total number of issues opened in their respective repositories each year that haven't been closed to date, grouped by associated language.
\end{itemize}

Using a combination of the above analytics, we computed the following metrics in Tableu:
\begin{itemize}
    \item \textbf{GH\_Popularity = } \\
    Mean(num\_repos, num\_users)
    \item \textbf{GH\_Availability = } \\
    Mean(num\_pull\_requests / num\_projects, num\_commits / num\_projects)
    \item \textbf{GH\_Demand = } \\
    Mean(num\_pending\_issues / num\_projects)
    \item \textbf{GH\_Community = } \\
    Mean(num\_commits / num\_projects, num\_projects/ num\_users, num\_commits / num\_users)
\end{itemize}

\subsection{StackOverflow}
With the StackOverflow data, we used the posts table described in Table \ref{so_posts} to compute the following intermediate metrics:
\begin{itemize}
  \item \textbf{num\_users - } total number of new users each year, grouped by associated language tag.
  \item \textbf{num\_questions - } total number of new questions asked each year, grouped by associated language tag.
  \item \textbf{num\_answers - } total number of answers to question each year, grouped by associated language tag.
  \item \textbf{total\_score - } total score of all posts each year, grouped by associated language tag.
  \item \textbf{num\_unanswered\_questions - } total number of questions asked each year that haven't received a response, grouped by associated language tag.
  \item \textbf{avg\_response\_time - } average time taken (in hours) for questions asked to receive a response each year, grouped by associated language tag.
\end{itemize}

Using a combination of the above analytics, we computed the following metrics in Tableu:
\begin{itemize}
    \item \textbf{SO\_Popularity = } \\
    Mean(num\_questions, num\_users)
    \item \textbf{SO\_Availability = } \\ 
    Mean(num\_answers / num\_questions)
    \item \textbf{SO\_Demand = } \\
    Mean(num\_unanswered\_questions / num\_questions)
    \item \textbf{SO\_Community = } \\ 
    Mean(avg\_response\_time, total\_score / num\_answers, num\_answers / num\_users, num\_questions / num\_users)
\end{itemize}

\subsection{Combined}
We combine the above metrics computed separately for Github and StackOverflow. We get the following combined metrics: 
\begin{itemize}
    \item $Popularity = $\\
    $0.5 * GH\_Popularity + (1 - 0.5) * SO\_Popularity$ \\
    \item $Availability = $ \\
    $0.5 * GH\_Availability + (1 - 0.5) * SO\_Availability$ \\
    \item $Demand = $ \\
    $0.5 * GH\_Demand + (1 - 0.5) * SO\_Demand$ \\
    \item $Community = $ \\ 
    $0.5 * GH\_Community + (1 - 0.5) * SO\_Community$ \\
\end{itemize}

Visualising these metrics provides valuable insights such as: the most popular programming languages, languages that have the most demand shortage, the languages with the most community engagement. It also highlights anomalies in the data that can be explained by real world events that happened in the past that had that effect. For example, we notice that swift rapidly increases in popularity while objective-c rapidly declines during 2013 - 2014. This could be explained due to the launch of swift in place of objective-c for iOS application development.

\section{Application Design}
In this section, we describe the overall architecture of our application as shown in Fig. \ref{fig:architecture}.

Our data sources are described in section IV. We ingest the data by downloading them into HDFS. The data is a raw dump and therefore we clean the data by dropping unwanted columns, handling null values and formatting special fields such as timestamps, etc. This is a very important step as it minimizes the possibility of run-time exceptions as well as significantly reduces the size of the data which speeds up processing. 

After cleaning, we profile the data to understand the characteristics of each column. We identify the data type and measure minimum, maximum as well as number of distinct values. This information is important as it helps us identify which columns can be joined as well as reason if that column needs to be reduced before a join.

Once we understand the data, we proceed to compute the analytic as described in the previous section. 

\begin{figure}[h]
\caption{User Interface design}
\centering
\label{fig:ui}
\includegraphics[width=0.48\textwidth]{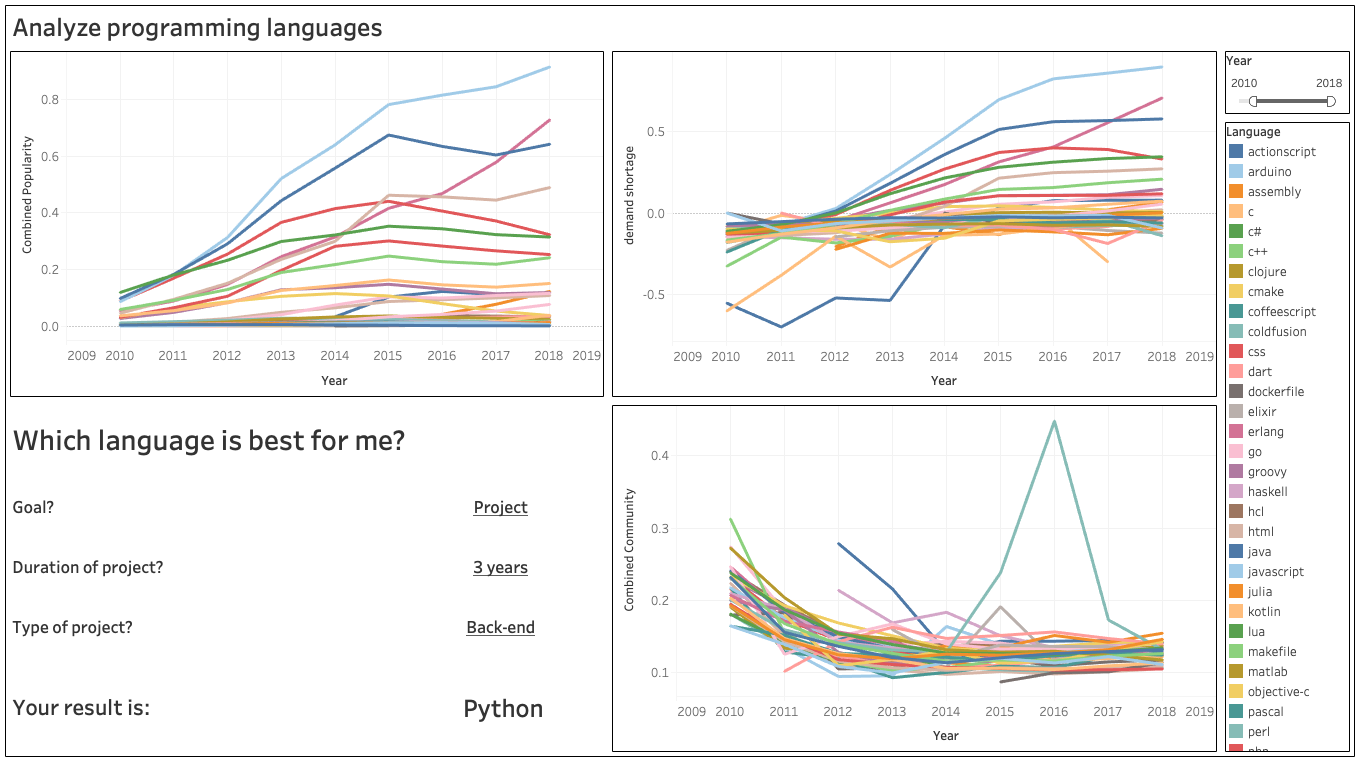}
\end{figure}

Once the processing is completed as illustrated in Fig. \ref{fig:architecture}, we obtain the analytic separately for each data source. Since the size of this data is quite small, we load it directly onto Tableu for further refinement and visualisation. We then visualize the results in a Tableu dashboard. The final design of our user interface is shown in Fig. \ref{fig:ui}. We take the goal of selecting the programming language from the user. If the goal is to build a project, we ask the user the duration of the project. If it is learning a language based on the demand shortage, we ask the time horizon of looking for a job. We ask the user the general category of the programming language of interest. Based on this information and based on the metrics we computed, we give the user a ranked list of recommendations of programming languages.

\section{Analysis}
We work with the two data sources separately at first. For both datasets we clean and join the relevant tables together. The major challenge is in joining the separate tables over a common column. This is because the data is huge and directly joining the tables results in the job getting killed due to exceeding the memory limit. For example, joining the commits and projects table over the $project\_id$ column results in the job getting killed because the commits table is huge and the result of the join exceeds memory limit. We tackled this problem by reducing the commits column by $year$ and $project\_id$ which significantly reduces the number of rows. Also, smart use of caching speeds up the process.

We also filter the entire dataset based on the top 50 most popular  programming languages found on GitHub.

After computing the intermediate metrics described in section V, we normalize the data using min-max normalisation procedure. We also stationarize the year on year data using first order differencing to remove the effect of increased usage of the Github and SO platforms over time.

We find the following insights based on the metrics computed and the vizualization:
\subsection{Popularity}

\begin{figure}[h]
\caption{Languages Legend}
\centering
\label{fig:legend}
\includegraphics[width=0.48\textwidth]{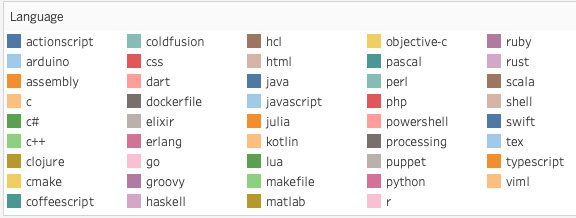}
\end{figure}

\begin{figure}[h]
\caption{Popularity}
\centering
\label{fig:popularity}
\includegraphics[width=0.48\textwidth]{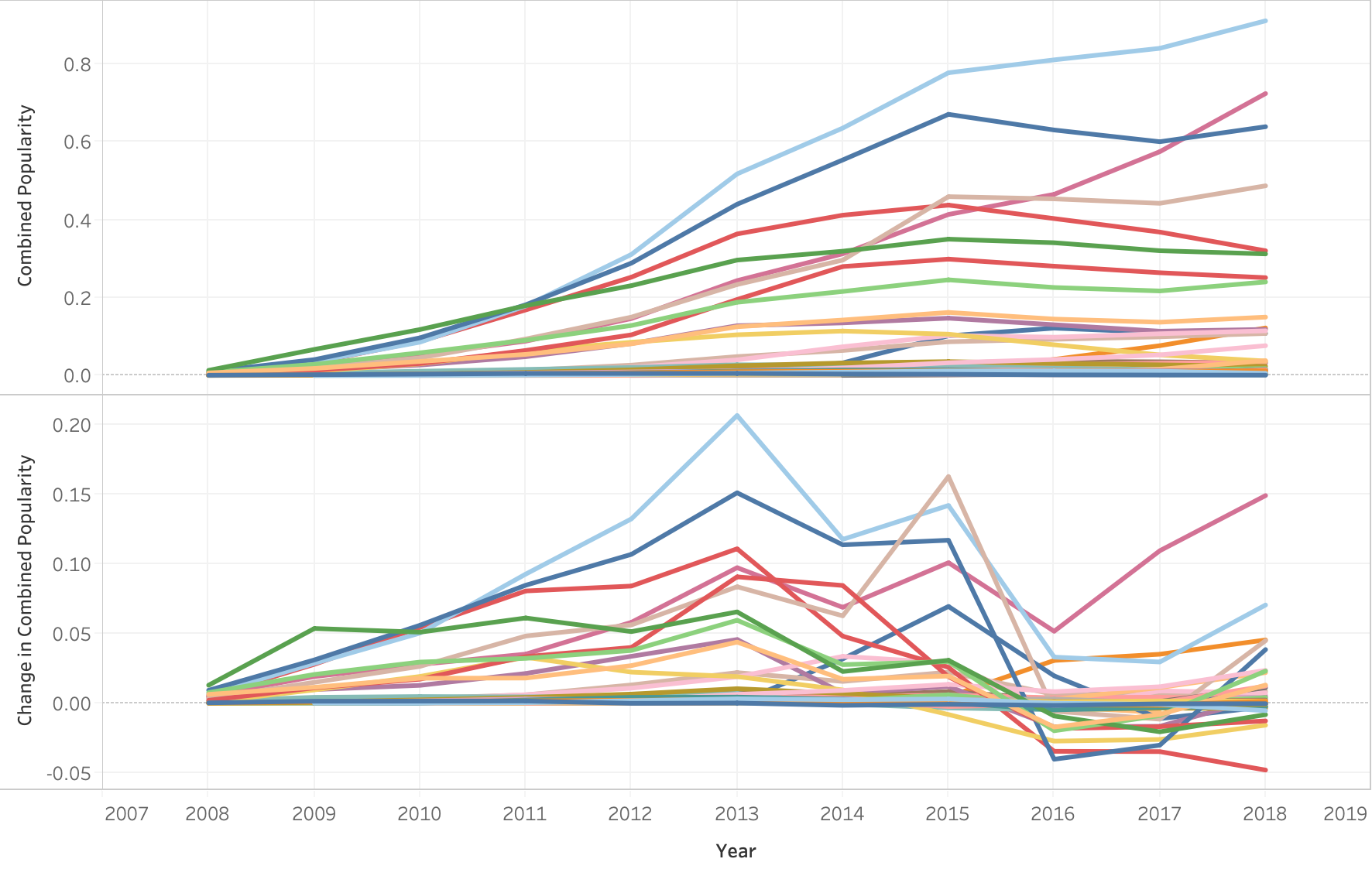}
\end{figure}

\begin{figure}[h]
\caption{Popular programming languages}
\centering
\label{fig:popularity_treemap}
\includegraphics[width=0.48\textwidth]{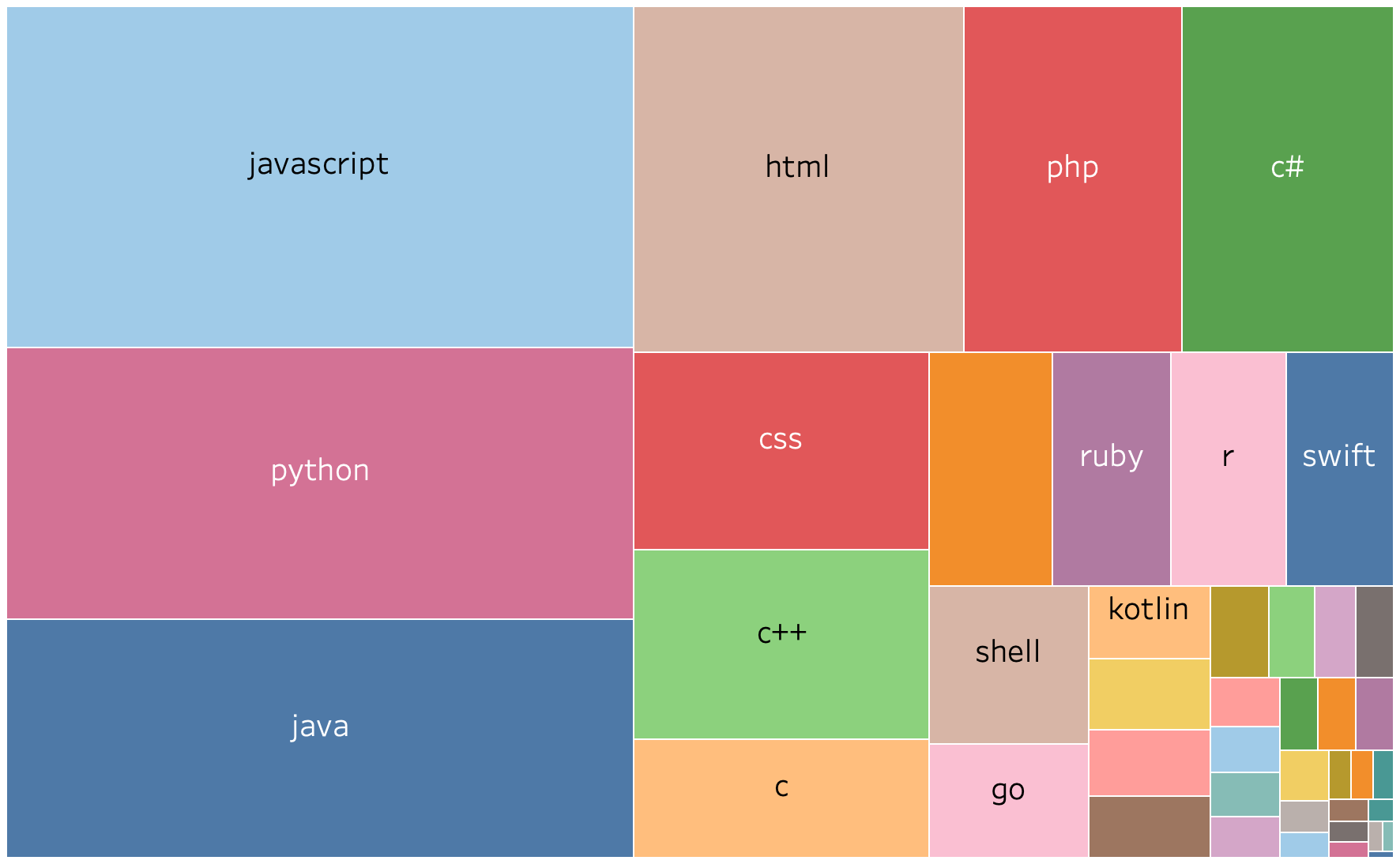}
\end{figure}

Unsurprisingly, Javascript and Python are the most popular languages, with Python overtaking Java in 2017.
Web Languages including Javascript, HTML, CSS along with Python are increasing the most in popularity. Go and Dart are the languages to watch out for.

\subsection{Demand - Availability}

\begin{figure}[h]
\caption{Demand Shortage}
\centering
\label{fig:demand_shortage}
\includegraphics[width=0.48\textwidth]{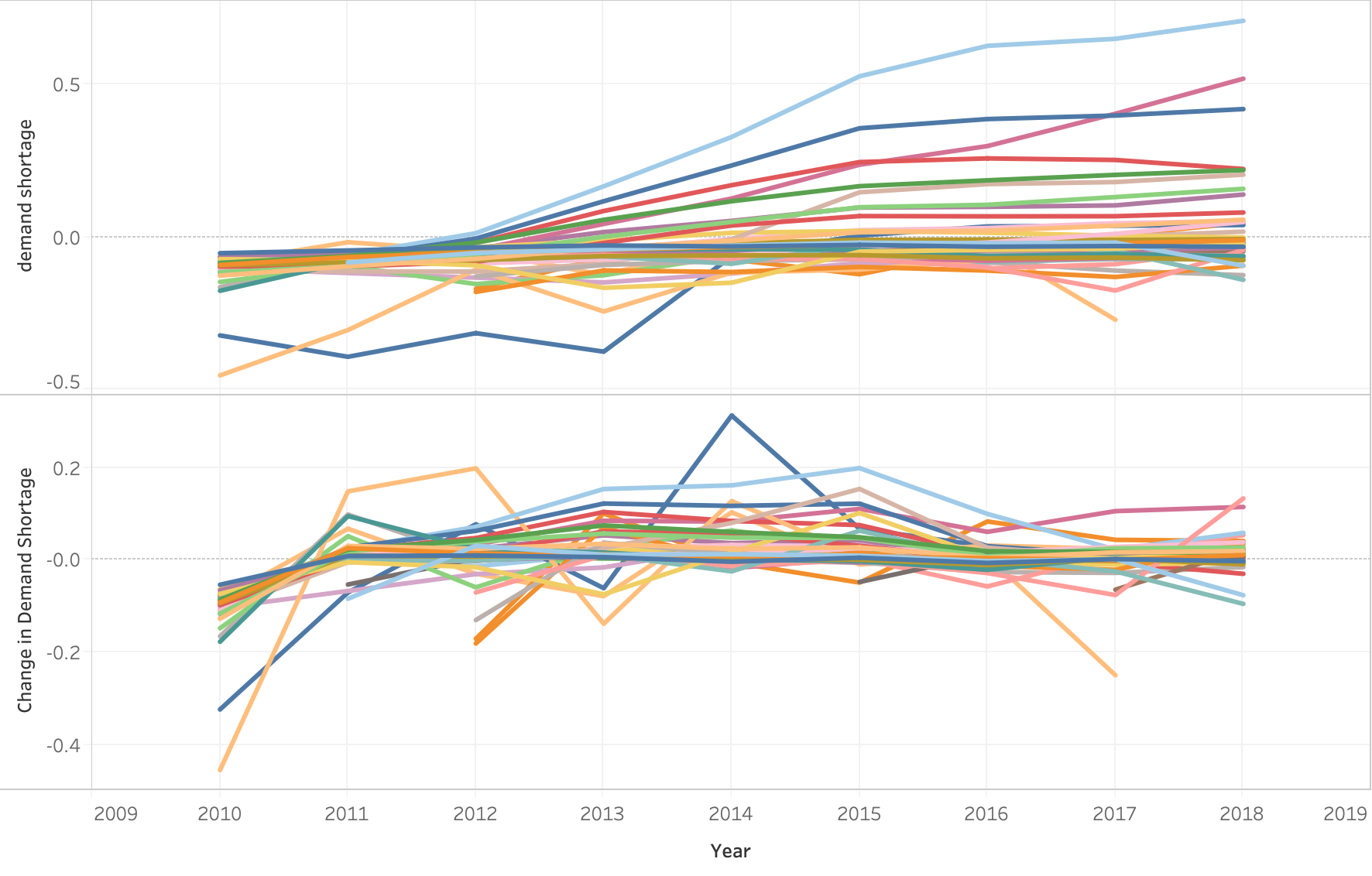}
\end{figure}

The most popular languages such as JS, Python and Java are also the ones which have the biggest demand shortage on SO and Github. 
Amongst the top are PHP and C\# whose popularity has been on a degrading over the last few years.
Go and Dart have the biggest rise in demand shortage, along with Python and JS.

\subsection{Community Engagement}

\begin{figure}[h]
\caption{Community Engagement}
\centering
\label{fig:community}
\includegraphics[width=0.48\textwidth]{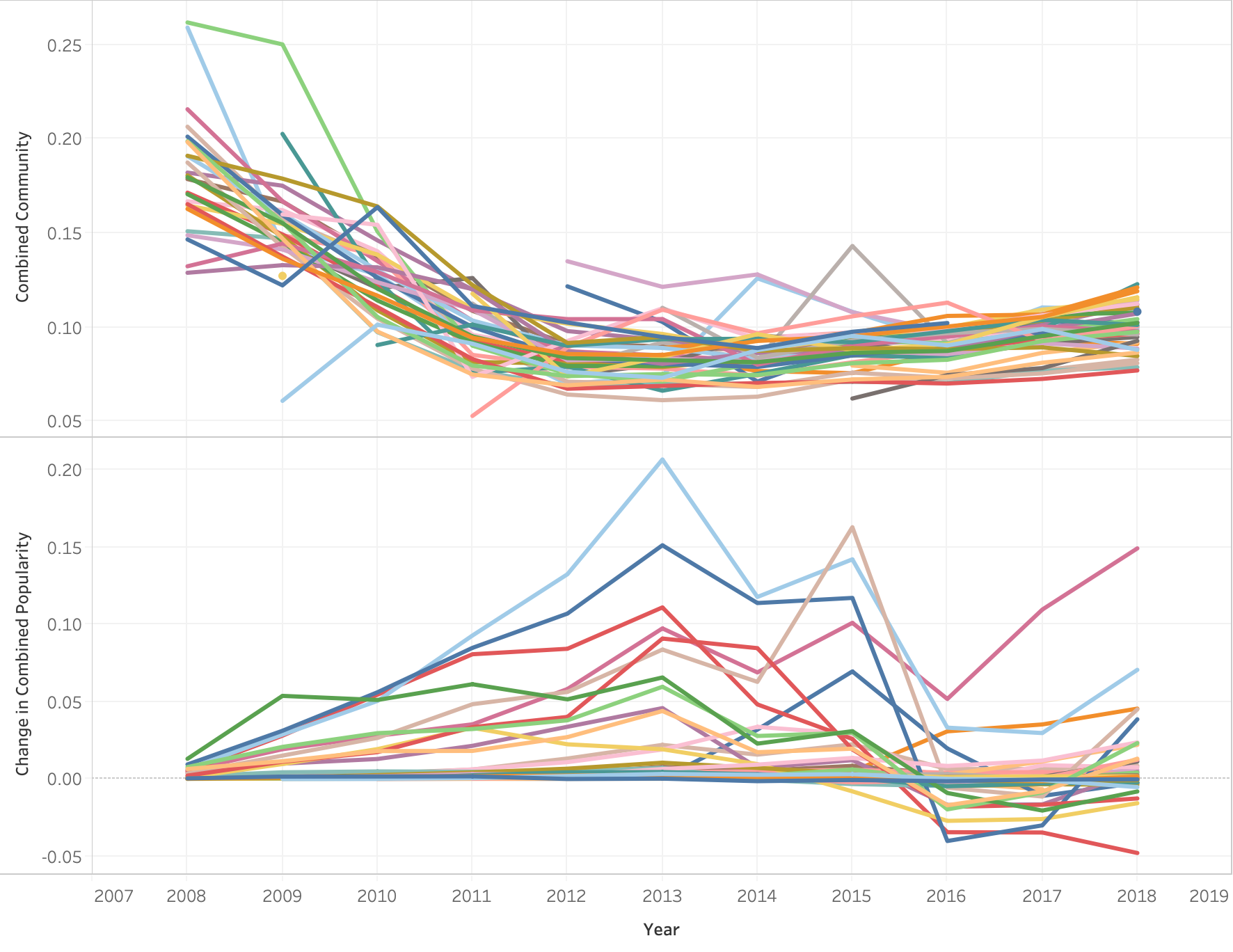}
\end{figure}

Ruby, Go, Julia, Dart and Rust have the most engaged online communities.

\section{Conclusion}
We analyse community characteristics of programming languages. In order to do so, we use data from Github and StackOverflow. We compute metrics intermediate metrics grouped by programming languages over the years. We use a combination of these metrics as a proxy for measures for qualities such as popularity, availability, demand and community engagement. We then visualise these qualities to analyse trends and provide relevant recommendation.

For students just starting to learn programming, exploring popular web based languages such as JavaScript along with HTML, CSS will be beneficial due to higher demand shortage as well as good community engagement. Investing time in learning Dart and Go which have the most engaged online communities as well as the highest rise in demand shortage will provide a competitive advantage. Languages on the decline in popularity such as PHP or C\# may not necessarily be a bad idea if looking for a job in the near future due to constant demand shortage.

For companies, working with the most popular languages gives easy access to talent, however, working with languages having high community engagement opens access to the top talent. Some of the best languages to consider are Ruby, Go, Julia, Dart and Rust.

\section{Future Work}
As part of the future work, the application could be extended to identify the type of the programming language. For example, C and Go could be categorised as "systems" as they are lower level and are well suited for systems development, Python could be categorised as back-end, etc. Then, the languages could only be analysed against languages belonging to the same category. This would enable a fairer comparison and would significantly improve the actuation of recommending languages or alternatives.

\section*{Acknowledgment}
Many thanks to Professor Suzanne McIntosh for the teaching and facilitation of the course. We also thank the NYU HPC team for the seamless provision of highly powerful infrastructure and Tableau for the free academic licence. 

\bibliographystyle{IEEEtran}
\bibliography{IEEEabrv, references}

\end{document}